\documentclass[10pt,twocolumn,twoside]{IEEEtran}
\usepackage{amsmath,amssymb}
\usepackage{latexsym}
\usepackage{psfrag}
\usepackage{epsfig}
\usepackage{graphicx,subfigure}
\usepackage{epstopdf}
\usepackage{multicol}
\usepackage{arydshln}
\usepackage{color}
\usepackage{bm}
\DeclareMathOperator{\sgn}{sgn}
 \DeclareMathOperator{\diag}{diag}
\newtheorem{theorem}{Theorem}
\newtheorem{proposition}{Proposition}

\newtheorem{definition}{Definition}
\newtheorem{lemma}{Lemma}
\newtheorem{remark}{Remark}

\newcommand{\R}{{\rm I\!R}}

\begin{document}
\title{\LARGE \bf Structural balance and opinion separation in\\ trust--mistrust social networks*}
\author{Weiguo Xia, Ming Cao, and Karl Henrik Johansson% <-this % stops a space
\thanks{*Compared to the journal version of this paper \cite{XiCaJo16}, an assumption has been added to both Theorem \ref{thm:5} and Theorem \ref{thm:7}.}
\thanks{This work was supported in part by grants from the European Research Council (ERC) under grant agreement ERC-StG-2012-307207, the Knut and Alice Wallenberg Foundation and the Swedish Research Council.}
\thanks{W. Xia is with the School of Control Science and Engineering, Dalian University of Technology, Dalian, 116024, China  (e-mail: wgxiaseu@dlut.edu.cn). M. Cao is with the Faculty of Mathematics and Natural Sciences, ENTEG, University of Groningen, The Netherlands (e-mail: m.cao@rug.nl). K. H. Johansson is with ACCESS Linnaeus Centre, School of Electrical Engineering, Royal Institute of Technology, Sweden.
 (e-mail: kallej@kth.se).}
}

%\markboth{IEEE Transactions on Control and Network Systems}
%{Xia \MakeLowercase{\textit{et al.}}: Structural balance and opinion separation}

\maketitle

\begin{abstract}
Structural balance theory has been developed in sociology and
psychology to explain how interacting agents, e.g., countries,
political parties, opinionated individuals, with mixed trust and
mistrust relationships evolve into polarized camps. Recent
results have shown that structural balance is necessary for
polarization in networks with fixed, strongly connected neighbor
relationships when the opinion dynamics are described by
DeGroot-type averaging rules. We develop this line of research in
this paper in two steps. First, we consider  fixed, \emph{not}
necessarily strongly connected, neighbor relationships. It is shown
that if the network includes a strongly connected subnetwork
containing mistrust, which influences the rest of the network, then
no opinion clustering is possible when that subnetwork is not
structurally balanced; all the opinions become neutralized in the
end. In contrast, it is shown that when that subnetwork is
indeed structurally balanced, the agents of the subnetwork evolve
into two polarized camps and the opinions of all other agents in
the network spread between these two polarized opinions. Second, we
consider \emph{time-varying} neighbor relationships. We show that
the opinion separation  criteria carry over if the conditions for
fixed graphs are extended to joint graphs. The results are developed for both discrete-time and continuous-time models.
\end{abstract}

\section{Introduction}

In theoretical sociology and social psychology, a strong interest has been maintained over the years in the study of the evolution of opinions of social groups \cite{Aj01,EaKl10}. There is a long tradition to study how continuous interactions within an interconnected collective without isolated subgroups, might lead to the emergence of segregation, or even polarization, of communities that form homogenous opinions only internally \cite{DuLi59,Sa74}. One popular theory is that the balance between trust and mistrust that dictate people's opinions to become closer or further apart, respectively, plays a major role in the dynamical process of opinion separation \cite{Fr06}. This theory, when explicitly expressed using signed graphs describing the trust and mistrust relationships among the interacting social agents, is  called \emph{structural balance theory} \cite{Ha53,CaHa56,Fe67,WaFa94,AbLu09}. Specifically, for the graph describing the neighbor relationships between agents in a social network, positive signs are assigned to those edges corresponding to trust and negative signs to those edges corresponding to mistrust. Then the network is \emph{structurally balanced} if all the vertices of its signed graph can be divided into two disjoint sets such that every edge between vertices in the same set is with a positive sign and every edge between vertices in the distinct sets is with a negative sign \cite{WaFa94}.

While structural balance theory tells clearly how the trust--mistrust relationships should be
distributed among the agents for the presence of stable polarized opinions, it does not specify how the agents'
opinions update. Recently, there is a growing effort to introduce DeGroot-type of opinion updating
rules to social networks with trust and mistrust relationships \cite{Al12,Al13,PrMaCa15,MeShJoCaHo15}.
 The DeGroot model \cite{De74} describes how each agent repeatedly updates its opinion to the
 average of those of its neighbors. Since this model reflects the fundamental human cognitive
 capability  of taking convex combinations when integrating related information \cite{An81}, it has been
 studied extensively in the past decades \cite{ChScLi13}.  But to show the process of opinion separation
  using the DeGroot model, more work \cite{HeKr02,Lo07,BlHeTs09} is to rely on mechanisms that lead to disconnected networks, the so-called bounded confidence Krause model, rather than to resort to
    trust--mistrust relationships in connected networks. Some other work has introduced
    an adaptive noisy updating model that characterizes individuals' diversified tendencies to explain
    the occurrence of clustering in human populations \cite{MaFlHe10}; in this model, noise is critical in sustaining clusters of opinions in a connected network.

For DeGroot-type opinion dynamics in trust--mistrust networks, it
has been proved in \cite{Al12,Al13} using continuous-time models
that in a strongly connected and structurally balanced network
consisting of two camps, where agents only trust those within the
same camp, the opinions of all the agents within the same camp
become the same, which is exactly the opposite of the opinion of the
other camp. It has also been shown that in a structurally unbalanced
network, the opinions of all the agents asymptotically converge to
zero. It remains an open question about how the opinions of the
agents evolve when the network is not strongly connected but
structurally unbalanced. What is even more intriguing is to
investigate the dynamical behavior under time-varying network
topologies, since in practical situations the relationships between
agents may change over time.

In this paper, we investigate the opinion evolution of interacting
agents with trust--mistrust relationships under either fixed network
topologies containing directed spanning trees or dynamically
changing topologies with joint connectivity. For the fixed topology
case, we show that when the network graph contains a strongly
connected subgraph with negative edges, which has a directed path to
every other vertex in the network, the opinions of all the agents
become neutralized at zero if the strongly connected subgraph is not
structurally balanced. In comparison, if the strongly connected
subgraph is structurally balanced, it is shown that the opinions of
the agents in this subgraph polarize at the exactly opposite values,
and the opinions of the rest of the agents lie in between the
polarized values. For dynamically changing network topologies,
similar conclusions hold when the graphical conditions are applied
to the corresponding joint graphs. Our results show
that in addition to getting polarized and reaching consensus, the
DeGroot-type opinion dynamics can give rise to opinion clustering in
a network under weaker connectivity conditions. This complements the
existing mechanisms that induce clusters in social networks through
introducing bounded confidence \cite{HeKr02}, updating noise
\cite{MaFlHe10}, or delays \cite{XiCa11}.

The rest of the paper is organized as follows. In Section~\ref{se:mot}, several examples are presented to motivate our study. In Section~\ref{se:problem}, we introduce the opinion dynamics models and formulate the problem considered in the paper. In Section~\ref{se:discrete} and Section~\ref{se:continuous} the  behaviors of the systems with discrete-time and continuous-time dynamics are studied, respectively. We present simulation examples  to verify the effectiveness of the theoretical results in Section~\ref{se:simulation}. Section~\ref{se:conclusion} discusses the conclusions and ideas for future work.

%=======================================================================
\section{Motivating example}\label{se:mot}

\begin{figure}
\begin{center} \includegraphics[width=3.2in]{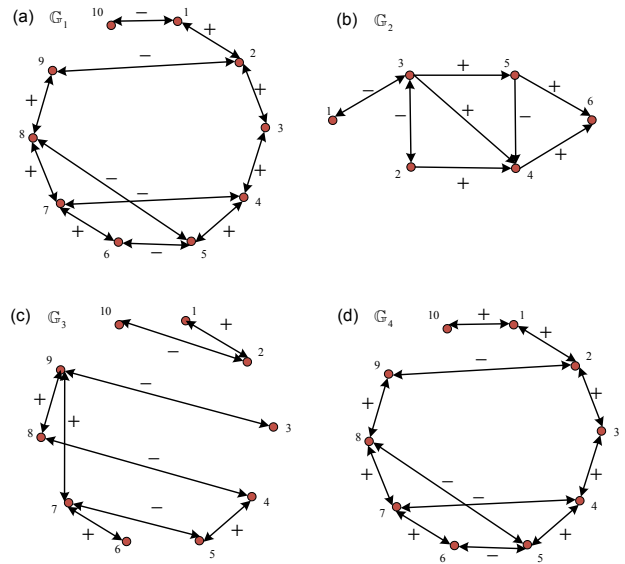}
\caption{Directed signed graphs $\mathbb{G}_1$, $\mathbb{G}_2$, $\mathbb{G}_3$ and $\mathbb{G}_4$.} \label{fig:ex1_1}
\end{center}
\end{figure}

In this section, to motivate introducing weaker connectivity
conditions for the DeGroot-type models in trust-mistrust social
networks, we present several examples showing that more complex
behaviors may take place compared to what have been reported in the
literature. Consider the directed graphs given in Fig.
\ref{fig:ex1_1}. Each directed edge is associated with a positive or
negative sign and the weight of each edge is either $1$ or $-1$.
Consider the network dynamics evolved on these graphs described
below. Each agent in the network is associated with a scalar $x_i\in
\R$ that represents its opinion on a certain subject. If $(v_j,v_i)$
is an edge in the graph, then agent $i$ takes agent $j$ as a
neighbor and thus agent $j$'s opinion is influencing agent $i$'s. Time is slotted. At each time instant, each agent updates
its state to the weighted average of its neighbors' and its own, and a positive weight 1 is assigned to its own opinion. Take
agent 2 in $\mathbb G_1$ in Fig. \ref{fig:ex1_1}(a) as an
example. Taking the weights of the edges into account, at time $t$,
the state of agent 2 is updated to
$$x_2(t+1)=\frac{x_1(t)+x_2(t)+x_3(t)-x_9(t)}{4},\ t\geq0,$$
since agents 1, 3, and 9 are the neighbors of agent 2 and $(v_9,v_2)$ with a negative weight gives the term $-x_9(t)$ in the equation. The other agents update their states in the same manner.

\begin{figure}
\begin{center} \includegraphics[width=3.2in]{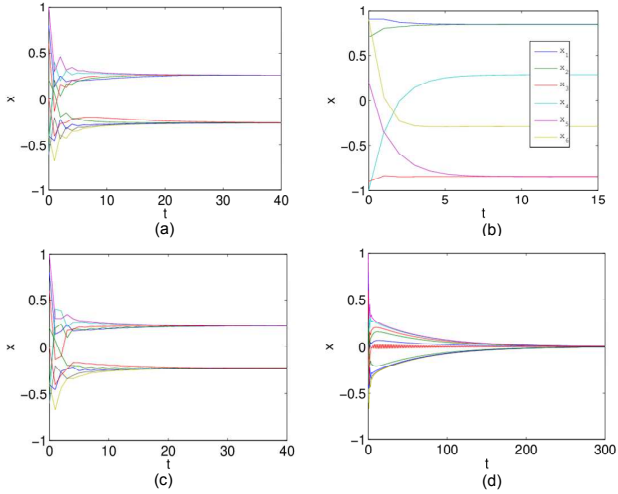}
\caption{The evolution of the agents' states when the graph topology (a) is $\mathbb{G}_1$; (b) is $\mathbb{G}_2$; (c) switches between $\mathbb{G}_1$ and $\mathbb{G}_3$; (d) switches between $\mathbb{G}_1$ and $\mathbb{G}_4$.} \label{fig:ex1_2}
\end{center}
\end{figure}

We are interested in the asymptotic behavior of the states of the
agents and  Fig. \ref{fig:ex1_2} shows  the evolution of the agents'
states under different network topologies: in (a) the graph is
$\mathbb{G}_1$; in (b) the graph  is $\mathbb{G}_2$; in (c) the
graph  is $\mathbb G_1$ at even times and is $\mathbb G_3$
otherwise; in (d) the graph  is $\mathbb G_1$ at even times and is
$\mathbb G_4$ otherwise. The initial condition of each agent lies in
the interval $[-1,1]$.

It is clear that $\mathbb G_1$ is structurally balanced \cite{Ha53}
(the formal definition will be given in the next section) in the
sense that it can be partitioned into $\{1,\ldots,5\}$ and $\{6,\ldots,{10}\}$
with positive edges within each set and negative edges in between.
Since $\mathbb G_1$ is strongly connected, we know from Theorem 1 in
\cite{Al13} that the agents' states will ``polarize" in the sense
that agents in $\mathcal V_1$ reach the same value that is opposite of the agreed value of those in $\mathcal V_2$. This is also called ``bipartite consensus" in \cite{Al13}. Although \cite{Al13}
only studied the switching case of strongly connected graphs at each
time, one may infer that the agents still polarize since $\mathbb
G_1$ and $\mathbb G_3$ are both structurally balanced and share the
same bipartition.

However, when the topology switches between $\mathbb G_1$ and
$\mathbb G_4$, it is unclear why the agents reach an agreement as
each graph is structurally balanced though they do not share a
common bipartition. What is intriguing is the phenomenon observed for
Fig. \ref{fig:ex1_2}(b) where the network topology $\mathbb G_2$
contains a directed spanning tree but is not strongly connected.
Instead of getting polarized or reaching consensus, the agents'
opinions become clustered and this clustering is a new behavior that
does not take place when the network is strongly connected. More
detailed theoretical analysis about such behavior is provided in
Theorem 2 and Theorem 5. The new opinion clustering phenomenon has
motivated us to study the system dynamics when the network
topologies are not strongly connected and/or become time varying.

\section{Problem formulation}\label{se:problem}

%\subsection{DeGroot-type opinion dynamics in trust--mistrust networks}
Consider a network of  $N$ agents labeled by $1,\ldots, N$, where each agent $i$, $1\leq i\leq N$, is associated with a scalar $x_i\in \R$ that represents its opinion on a certain subject. Here, $x_i$ being positive implies supportive opinions, being negative implies protesting views, and being zero implies neutral reaction. We use a directed signed graph $\mathbb G$ \cite{CaHa56} with the vertex set $\mathcal V = \{v_1, \ldots, v_N\}$ to describe the trust--mistrust relationships between the agents. The definition of signed graphs is as follows.
\begin{definition}\label{def:signed}
A directed {\it signed} graph is a directed graph in which each edge is associated with either a positive or negative sign.
\end{definition}

Some notions in graph theory need to be introduced \cite{BaGu00}. We consider only directed graphs without self-loops throughout the paper. In a directed graph $\mathbb G=(\mathcal V,\mathcal E)$ with $\mathcal V=\{v_1,\ldots,v_N\}$ and $\mathcal E\subseteq\mathcal V\times \mathcal V$, a directed \emph{walk} is a sequence of vertices $v_{i_1},\ldots,v_{i_k}$ such that $(v_{i_s},v_{i_{s+1}})\in\mathcal{E}$ for $s=1,\ldots,k-1$. A directed \emph{path} is a walk with distinct vertices in the sequence. A directed \emph{cycle} is a walk with distinct vertices $v_{i_1},\ldots,v_{i_k}$, $k\geq2$ and $v_{i_1}=v_{i_k}$. $\mathbb{G}$ is said to be \textit{strongly connected} if there is a directed path from every vertex to every other vertex in $\mathbb{G}$. A directed \emph{tree} is a graph containing a unique vertex, called \emph{root}, which has a directed path to every other vertex. A directed \emph{spanning tree} $\mathbb G^s=(\mathcal V^s,\mathcal E^s)$ of the directed graph $\mathbb G=(\mathcal V,\mathcal E)$ is a subgraph of $\mathbb G$ such that $\mathbb G^s$ is a directed tree and $\mathcal V^s=\mathcal V.$ $\mathbb G$ is said to contain a directed spanning tree if a directed spanning tree is a subgraph of $\mathbb G$. ``Directed" is omitted for the rest of the paper and we simply say $\mathbb G$ contains a spanning tree since we focus exclusively on directed graphs. The \emph{root vertex set} of $\mathbb G$ is a set of all the roots of $\mathbb G$.

In the $N$-agent network, there is a directed edge $(v_j,v_i)$ from $v_j$ to $v_i$ if and only if agent $i$ takes agent $j$ as a neighbor and thus agent $j$'s opinion is influencing agent $i$'s. Furthermore,  the directed edge $(v_j,v_i)$ is assigned with a non-zero weight $a_{ij}$, which is positive if agent $i$ trusts agent $j$  and negative otherwise; here, we assume the inter-agent relationship, if there is any, is either trusting or mistrusting, although the strength of the relationship may vary as reflected by the magnitude $|a_{ij}|$.  We use $\mathcal N_i$ to denote the set of indices of agent $i$'s neighbors.

For the DeGroot-type updating rule, both discrete-time and continuous-time models have been constructed in the literature. The discrete-time opinion dynamics can be described by
\begin{equation} \label{eq:system1_1}
x_i(t+1)=\sum_{j\in \{\mathcal N_i (t),i\}}p_{ij}(t)x_j(t), \ i= 1,\ldots, N, t=0, 1, \ldots,
\end{equation}
where $a_{ii}(t)>0$ is a self-trusting weight and
\begin{equation}
p_{ij}(t) = \frac{a_{ij}(t)}{ a_{ii}(t)+ \sum_{k\in \mathcal N_i (t)}  |a_{ik}(t)|},
\end{equation}
which obviously satisfies
\begin{equation}\label{eq:rowSumT}
p_{ii}(t) + \sum_{j\in \mathcal N_i (t)} |p_{ij}(t)| =1.
\end{equation}
If we take $\ x(t)=[x_1(t),\ldots,x_N(t)]^T$ to be the network state, then equation (\ref{eq:system1_1}) can be written in its state-space form
\begin{equation} \label{eq:system1_2}
x(t+1)=P(t)x(t), \qquad t=0, 1, \ldots,
\end{equation}
where $P(t)=(p_{ij}(t))_{N\times N}$ is an $N\times N$ matrix with positive diagonals.

Similarly, the continuous-time update equation for agent $i$ is
\begin{equation} \label{eq:system2_1}
\dot{x}_i=-\sum_{j\in \mathcal{N}_i(t)}|a_{ij}(t)|\bigg(x_i-\sgn(a_{ij}(t))x_j\bigg), \ i=1, \ldots,N,
\end{equation}
where $\sgn(\cdot)$ denotes the sign function. System (\ref{eq:system2_1}) can be written in the compact form
\begin{equation} \label{eq:system2_2}
\dot{x}=-L(t)x,
\end{equation}
where $L(t)$ is the signed Laplacian matrix that is defined by
\begin{align}\label{eq:signedL}
&l_{ii}(t)=\sum_{j\in \mathcal N_i(t)}|a_{ij}(t)|,\nonumber\\
&l_{ij}(t)=\left\{
\begin{array}{ll}
-a_{ij}(t), & \textrm{ for }\ j\in \mathcal{N}_i(t),\\
0, & \begin{matrix}
      \textrm{ for } j\neq i \textrm{ and }  j\notin \mathcal{N}_i(t). \\
    \end{matrix}
\end{array}
\right.
\end{align}

Since the graphs describing the interactions between agents may change with time $t$, we use $\mathbb G(P(t))$ and $\mathbb G(L(t))$ to denote the graph at time $t$ for the discrete-time system (\ref{eq:system1_2}) and for the continuous-time system (\ref{eq:system2_2}), respectively. Let $P_{11}(t)$ be the principal submatrix of $P(t)$ obtained from $P(t)$ by deleting the $i_{1}$-th, $i_{2}$-th,$\ldots$, $i_m$-th rows and columns, where $1\leq i_1,\ldots,i_m\leq N$. Then $\mathbb G(P_{11}(t))=(\mathcal V^s,\mathcal E^s)$ denotes the subgraph of $\mathbb G(P(t))=(\mathcal V,\mathcal E)$ such that $\mathcal V^s=\mathcal V\backslash\{v_{i_1},\ldots,v_{i_m}\}$ and $\mathcal E^s=\{(v_i,v_j):\ (v_i,v_j)\in\mathcal E\ \text{and }v_i,v_j\in\mathcal V^s\}$. When the graph is fixed, we omit $t$ and write $\mathbb G(P)$ and $\mathbb G(L)$.

Equations (\ref{eq:system1_2}) and (\ref{eq:system2_2}) both come from the DeGroot model and share the same intuition: the opinions of those agents that agent $i$ trusts influence its opinion positively and thus the averaging rule tries to bring them closer; at the same time, the opinions of those agents that agent $i$ does not trust influence its opinion negatively and thus the averaging rule pushes them apart. It is natural then that the distribution of positive and negative edges of a graph affects the evolution of opinions and for this reason, the notion of ``structural balance" becomes instrumental.

\begin{definition}\label{def:balanced}
A directed signed graph $\mathbb{G}$ with vertex set $\mathcal{V}$ is {\it structurally balanced} if $\mathcal{V}$ can be partitioned into two disjoint subsets $\mathcal{V}_1$ and $\mathcal{V}_2$ such that all the edges $(v_i,v_j)$ with $v_i,v_j$ taken in the same set $\mathcal V_k$, $k=1, 2$, are of positive signs and all the edges $(v_i,v_j)$ with $v_i,v_j$ taken in different sets $\mathcal V_k$ are of negative signs.
\end{definition}

Note that in the definition of structural balance, a network is still said to be structurally balanced if \emph{all} the edges of the network graph are assigned with the positive sign and thus one of $\mathcal V_1$ and $\mathcal V_2$ in Definition \ref{def:balanced} becomes empty.

A social network that is structurally balanced in which the agents' opinions update according to the DeGroot-type  averaging rules (\ref{eq:system1_1}) or (\ref{eq:system2_1})  may evolve into two polarized camps. We now make our notion of polarization precise.

\begin{definition}\label{def:biConsensus}
System (\ref{eq:system1_2}) or  (\ref{eq:system2_2}) \emph{polarizes} if for almost all initial conditions, $\lim_{t\rightarrow\infty}|x_i(t)|=\lim_{t\rightarrow\infty}|x_j(t)|>0$ for all $i,j=1,\ldots,N$, and $\lim_{t\rightarrow\infty}x_i(t)=-\lim_{t\rightarrow\infty}x_j(t)$ for some $i\neq j$.
\end{definition}

It has been shown in \cite{Al13} that system  (\ref{eq:system2_2}) with a fixed, strongly connected network graph $\mathbb G(L)$ polarizes if  $\mathbb G(L)$ is structurally balanced. It is the goal of this paper to study for both systems (\ref{eq:system1_2}) and  (\ref{eq:system2_2}), what the relationship is between structural balance and opinion separation, for which opinion polarization is an extreme case, when the network topology is either fixed and contains a spanning tree or time varying. In what follows, we study separately the discrete-time model  (\ref{eq:system1_2}) and the continuous-time model (\ref{eq:system2_2}).

%=======================================================================
\section{Discrete-time model}\label{se:discrete}

We introduce a $2N$-dimensional system\footnote{We are indebted to Julien Hendrickx for pointing out this reformulation of the update equations.} \cite{He14} based on the $N$-dimensional system (\ref{eq:system1_2}). For a matrix $P(t)$, define two nonnegative matrices $P^+(t)$ and $P^-(t)$ according to
\begin{equation}\label{eq:P}
\begin{split}
&(P^+(t))_{ij}=\begin{cases}
p_{ij}(t), &p_{ij}(t)>0,\\
0, &p_{ij}(t)\leq0,
\end{cases}\\
&(P^-(t))_{ij}=\begin{cases}
-p_{ij}(t), &p_{ij}(t)<0,\\
0, &p_{ij}(t)\geq0,
\end{cases}
\end{split}
\end{equation}
where $(P^+(t))_{ij}$ and $(P^-(t))_{ij}$ are the $ij$-th elements of $P^+(t)$ and $P^-(t)$, respectively. It is obvious that $P(t)=P^+(t)-P^-(t)$. Define $x^+_i(t)=x_i(t)$, $x^-_i(t)=-x_i(t)$. One knows that $x^+_i(t)+x^-_i(t)=0$ for all $t\geq0$. From system (\ref{eq:system1_2}), we obtain the following update equations for $x^+_i(t)$ and $x^-_i(t)$:
\begin{align}\label{eq:ysystem1}
x^+_i(t+1)=\sum_{j,\ p_{ij}(t)>0}p_{ij}(t)x^+_j(t)+\sum_{j,\ p_{ij}(t)<0}|p_{ij}(t)|x^-_j(t),\nonumber\\
x^-_i(t+1)=\sum_{j,\ p_{ij}(t)>0}p_{ij}(t)x^-_j(t)+\sum_{j,\ p_{ij}(t)<0}|p_{ij}(t)|x^+_j(t).
\end{align}
Let $y(t)=[x^+_1(t),\ldots,x^+_N(t),x^-_1(t),\ldots,x^-_N(t)]^T.$ Then system (\ref{eq:ysystem1}) can be written as
\begin{equation}\label{eq:ysystem2}
y(t+1)=\begin{bmatrix}
      P^+(t) & P^-(t)  \\
      P^-(t)  & P^+(t)  \\
    \end{bmatrix}y(t)=Q(t)y(t),
\end{equation}
where $Q(t)=\begin{bmatrix}
      P^+(t) & P^-(t) \\
      P^-(t) & P^+(t) \\
    \end{bmatrix}$ is a stochastic matrix.

To study the properties of system (\ref{eq:system1_2}), we will explore the properties of system (\ref{eq:ysystem2}) which is a classical consensus system and existing convergence results can be utilized \cite{ReBe05,CaMoAn08a,HeTs13}. We make the connection between the  graph $\mathbb{G}(P(t))$ associated with $P(t)$ with the vertex set $\{v_1,\ldots,v_N\}$ and the graph $\mathbb{G}(Q(t))$ associated with $Q(t)$ with the vertex set $\{v_1^\prime,v_2^\prime,\ldots,v_{2N}^\prime\}$, so that we can transform the graphical conditions on $\mathbb{G}(P(t))$ to conditions on $\mathbb{G}(Q(t))$.

\begin{figure}
\begin{center} \includegraphics[width=2.3in]{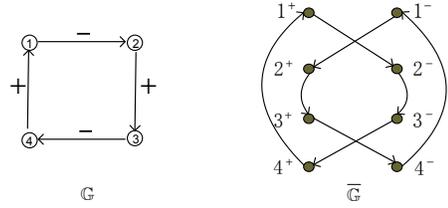}
\caption{A structurally balanced graph $\mathbb{G}$ and the corresponding graph $\bar{\mathbb{G}}$.} \label{fig:ex2_1}
\end{center}
\end{figure}

\begin{figure}
\begin{center} \includegraphics[width=2.3in]{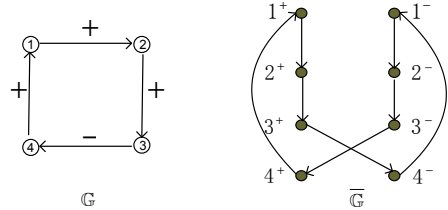}
\caption{A structurally unbalanced graph $\mathbb{G}$ and the corresponding graph $\bar{\mathbb{G}}$.} \label{fig:ex2_2}
\end{center}
\end{figure}

Given a directed signed graph $\mathbb{G}=(\mathcal{V},\mathcal{E})$ with $\mathcal{V}=\{v_1,v_2,\ldots,v_N\}$, we define an  enlarged directed graph $\bar{\mathbb{G}}=(\bar{\mathcal{V}},\bar{\mathcal{E}})$ based on $\mathbb{G}$ as follows. $\bar{\mathbb{G}}$ has $2N$ vertices and all of its edges are positive. Denote the vertices in $\bar{\mathbb{G}}$ as $v_1^+,v_1^-,v_2^+,v_2^-,\ldots,v_N^+,v_N^-$. If there is a positive edge $(v_i,v_j)\in\mathcal{E}$, then there are two directed edges $(v_i^+,v_j^+),(v_i^-,v_j^-)\in\bar{\mathcal{E}}$; if there is a negative edge $(v_i,v_j)\in\mathcal{E}$, then there are two directed edges $(v_i^+,v_j^-),(v_i^-,v_j^+)\in\bar{\mathcal{E}}$. In this manner, it is easy to see that if there is a positive path\footnote{In a  directed signed graph, a path is said to be positive if it contains an even number of edges with negative weights and to be negative otherwise. A positive or negative cycle is defined similarly.} from $v_i$ to $v_j$ in $\mathbb{G}$, then there is a path from $v_i^+$ to $v_j^+$ and a path from $v_i^-$ to $v_j^-$ in $\bar{\mathbb{G}}$; if there is a negative path from $v_i$ to $v_j$ in $\mathbb{G}$, then there is a path from $v_i^+$ to $v_j^-$ and a path from $v_i^-$ to $v_j^+$ in $\bar{\mathbb{G}}$. Two examples are illustrated in Figs.~\ref{fig:ex2_1} and \ref{fig:ex2_2}.

\begin{lemma}\label{lm:7}
Let $\mathbb{G}$ be a strongly connected signed graph and let $\bar{\mathbb{G}}$ be the enlarged graph based on $\mathbb{G}$. Then $\mathbb{G}$ is structurally balanced if and only if $\bar{\mathbb{G}}$ is disconnected and composed of two strongly connected components.
\end{lemma}
{\it Proof.} (Necessity) If  $\mathbb{G}$ is structurally balanced, there is a bipartition $\mathcal{V}_1=\{v_{i_1},\ldots,v_{i_m}\},\ \mathcal{V}_2=\{v_{i_{m+1}},\ldots,v_{i_N}\}$, $1\leq i_1,\ldots,i_N\leq N$ of $\mathcal{V}$ such that the edges between $\mathcal{V}_1$ and $\mathcal{V}_2$ are all negative and edges within each set $\mathcal{V}_i,\ i=1,2,$ are all positive. We claim that in $\bar{\mathbb{G}}$ there is no edge between $\bar{\mathcal{V}}_1=\{v^+_{i_1},\ldots,v^+_{i_m},v^-_{i_{m+1}},\ldots,v^-_{i_N}\}$ and $\bar{\mathcal{V}}_2=\{v^-_{i_1},\ldots,v^-_{i_m},v^+_{i_{m+1}},\ldots,v^+_{i_N}\}$, and thus $\bar{\mathbb{G}}$ is disconnected. If the contrary is true and there is an edge between $\bar{\mathcal{V}}_1$ and $\bar{\mathcal{V}}_2$, then there is a positive edge between $\mathcal{V}_1$ and $\mathcal{V}_2$ or a negative edge within $\mathcal{V}_1$ or $\mathcal{V}_2$, which contradicts the fact that $\mathbb{G}$ is structurally balanced. Since $\mathbb{G}$ is strongly connected, each component with the vertex set  $\bar{\mathcal{V}}_i,\ i=1,2,$ is strongly connected.

(Sufficiency) Assume that the vertex sets of the two components are $\bar{\mathcal V}_1$ and $\bar{\mathcal V}_2$ and $v^+_{i_1},\ldots,v^+_{i_m}$ are in $\bar{\mathcal V}_1$ and $v^+_{i_{m+1}},\ldots,v^+_{i_N}$ are in $\bar{\mathcal V}_2$. We claim that $v^-_{i_j}, j=1,\ldots,m$ are in $\bar{\mathcal V}_2$ and $v^-_{i_j}, j=m+1,\ldots,N$ are in $\bar{\mathcal V}_1$. If this is not true, then without loss of generality, assume $v^+_{i_1}$ and $v^-_{i_1}$ are both in $\bar{\mathcal V}_1$. Since each component is strongly connected, there is a path from $v_{i_1}^+$ to $v_{i_j}^+$ and a path from $v_{i_j}^+$ to $v_{i_1}^-$ for $j=2,\ldots,m$. Then from the definition of $\bar{\mathbb{G}}$, there is a path from $v_{i_1}^-$ to $v_{i_j}^-$ and a path from $v_{i_j}^-$ to $v_{i_1}^+$. It follows that $v_{i_j}^-,\ j=2,\ldots,m$ are in $\bar{\mathcal V}_1$ and thus  $\bar{\mathcal V}_1=\{v^+_{i_j},v^-_{i_j},j=1,\ldots,m\}$. Similarly $\bar{\mathcal V}_2=\{v^+_{i_j},v^-_{i_j},j=m+1,\ldots,N\}$. Since there is no edge between $\bar{\mathcal V}_1$ and $\bar{\mathcal V}_2$ in $\bar{\mathbb{G}}$, it follows that $\mathbb{G}$ is disconnected, which contradicts the assumption. One can conclude that  $\bar{\mathcal{V}}_1=\{v^+_{i_1},\ldots,v^+_{i_m},v^-_{i_{m+1}},\ldots,v^-_{i_N}\}$ and $\bar{\mathcal{V}}_2=\{v^-_{i_1},\ldots,v^-_{i_m},v^+_{i_{m+1}},\ldots,v^+_{i_N}\}$. Then it is easy to see that $\mathbb{G}$ is structurally balanced if we define $\mathcal{V}_1=\{v_{i_1},\ldots,v_{i_m}\},\ \mathcal{V}_2=\{v_{i_{m+1}},\ldots,v_{i_N}\}$. \hfill $\Box$

\begin{lemma}\label{lm:8}
Let $\mathbb{G}$ be a strongly connected signed graph and let $\bar{\mathbb{G}}$ the enlarged graph based on $\mathbb{G}$. Then $\mathbb{G}$ is structurally unbalanced if and only if $\bar{\mathbb{G}}$ is strongly connected.
\end{lemma}
\noindent{\it Proof.} Sufficiency is obvious in view of Lemma \ref{lm:7} and we only prove the necessity.

Since $\mathbb{G}$ is structurally unbalanced, without loss of generality, assume that there is a negative cycle from $v_k$ to $v_k$ in $\mathbb{G}$ \cite{CaHa56}. For any $i\neq j,\ 1\leq i,j\leq N$, since $\mathbb{G}$ is strongly connected, there is a path from $v_i$ to $v_j$. Without loss of generality, assume this path is positive. Then there is a directed path from $v_i^+$ to $v_j^+$ and a directed path from $v_i^-$ to $v_j^-$ in $\bar{\mathbb{G}}$. Since $\mathbb{G}$ is strongly connected and there is a negative cycle starting from $v_k$, we are able to find a directed negative walk from $v_j$ to $v_i$. Accordingly, there is a walk from $v_j^+$ to $v_i^-$ in $\bar{\mathbb{G}}$ . Thus in $\bar{\mathbb{G}}$ there is a directed walk from $v_i^+$ to $v_j^+$, to  $v_j^-$ and to $v_i^-$. Thus  there is a directed path from $v_i^+$ to $v_j^+$, to  $v_j^-$ and to $v_i^-$ and it follows that $\bar{\mathbb{G}}$ is strongly connected.
\hfill $\Box$

Two examples are given in Figs. \ref{fig:ex2_1} and \ref{fig:ex2_2} to illustrate these two lemmas.

Let $\overline{\mathbb{G}(P(t))}$ be the enlarged graph based on $\mathbb{G}(P(t))$. If we denote the graph associated with $Q(t)$ by $\mathbb{G}(Q(t))$, then it is easy to see from the structure of $Q(t)$ that $\overline{\mathbb{G}(P(t))}$ and $\mathbb{G}(Q(t))$ are isomorphic \cite{BaGu00}, i.e.,  $\overline{\mathbb{G}(P(t))}\simeq\mathbb{G}(Q(t))$. The bijection $\phi$ that maps the vertex set $\{v_1^+,v_1^-,\ldots,v_N^+,v_N^-\}$ of $\overline{\mathbb{G}(P(t))}$ to the vertex set $\{v_1^\prime,v_2^\prime,\ldots,v_{2N}^\prime\}$ of $\mathbb{G}(Q(t))$ is given by: $v_i^+\rightarrow v_i^\prime,\ v_i^-\rightarrow v^\prime_{N+i}$ for $i=1,\ldots,N$. We simply use $\{v_1^+,v_1^-,\ldots,v_N^+,v_N^-\}$ to denote the vertex set for $\mathbb{G}(Q(t))$ in the following for the clear correspondence. We have the following result from Lemmas \ref{lm:7} and \ref{lm:8}.

\begin{proposition}\label{re:1}
Assume that $\mathbb{G}(P(t))$ is strongly connected. $\mathbb{G}(P(t))$ is structurally balanced if and only if $\mathbb{G}(Q(t))$ is disconnected and composed of two strongly connected components; $\mathbb{G}(P(t))$ is structurally unbalanced if and only if $\mathbb{G}(Q(t))$ is strongly connected.
\end{proposition}

\subsection{$\mathbb G(P(t))$ is fixed}

When $\mathbb G(P(t))$ is fixed, the matrix $P(t)$ in (\ref{eq:system1_2}) and $Q(t)$ in  (\ref{eq:ysystem2}) are also fixed, i.e., $P(t)\equiv P,\ Q(t)\equiv Q,\ t\geq0$. In view of (\ref{eq:rowSumT}), we know that\footnote{We take the absolute value of a matrix elementwise.} $|P|$ is a stochastic matrix.

\begin{theorem}\label{thm:1}
Consider an irreducible $P$ with $|P|$ being stochastic. Assume that the graph $\mathbb{G}(P)$ has at least one negative edge. System (\ref{eq:system1_2}) polarizes if and only if $\mathbb{G}(P)$ is structurally balanced. If $\mathbb{G}(P)$ is structurally unbalanced, then $\lim_{t\rightarrow\infty}x(t)=0$ for every initial value.
\end{theorem}
{\it Proof:} (Sufficiency) When $\mathbb{G}(P)$ is structurally balanced with at least one negative edge, from the proof of Lemma \ref{lm:7} and Proposition \ref{re:1} we know that $\mathbb{G}(Q)$ contains two disconnected components each of which is strongly connected and $\{v_{i_1}^+,\ldots,v_{i_{m}}^+,v_{i_{m+1}}^-,\ldots,v_{i_N}^-\}$ and $\{v_{i_1}^-,\ldots,v_{i_{m}}^-,v_{i_{m+1}}^+,\ldots,v_{i_N}^+\}$ are the vertex sets of the two components, for some $m$, $1\leq m<N$ and $1\leq i_1,\ldots,i_N\leq N$. Thus the $y$-system (\ref{eq:ysystem2}) is decomposed into two disconnected subsystems. It follows from the classical consensus result \cite{ReBe05,CaMoAn08a} that each subsystem converges to some constant value. In system (\ref{eq:system1_2}) the agents in $\{v_{i_1},\ldots,v_{i_{m}}\}$ have the same value and the other agents in $\{v_{i_{m+1}},\ldots,v_{i_{N}}\}$ have the opposite value. Since the initial conditions that renders the agreed value of each component to be zero form a set which has zero Lebesgue measure, system (\ref{eq:system1_2}) polarizes.

(Necessity) Assume that system (\ref{eq:system1_2}) polarizes. If $\mathbb{G}(P)$ is structurally unbalanced, then $\mathbb{G}(Q)$ is strongly connected based on Proposition \ref{re:1}. It follows that $y_i(t)$ converges to some constant $\alpha$ for all $i=1,\ldots,2N$ as $t\rightarrow\infty$. Since the $y$-system contains $x_i^+(t)$ and $x^-_i(t)$ as subsystems, $\alpha$ should always be 0 which contradicts the assumption that system (\ref{eq:system1_2}) polarizes. One can conclude that $\mathbb{G}(P)$ is structurally balanced.
 \hfill $\Box$

The above discussion has assumed that $P$ is irreducible or equivalently $\mathbb{G}(P)$ is strongly connected. Next we discuss the more general case when $\mathbb{G}(P)$ is not necessarily strongly connected but contains a spanning tree. Using some permutation of rows and columns of $P$, $P$ can be transformed into
\begin{equation}\label{eq:Ptree}
P=\begin{bmatrix}
    P_{11} & \bm{0} \\
    P_{21} & P_{22} \\
  \end{bmatrix}
  \end{equation}
where $P_{11}\in\R^{r\times r},\ P_{22}\in\R^{(N-r)\times(N-r)}$, $P_{21}\in\R^{(N-r)\times r}$ and $\bm{0}$ is a zero matrix of compatible dimension. The submatrix $P_{11}$ is irreducible and the subgraph $\mathbb{G}(P_{11})$ is strongly connected, and there is a directed path from every vertex in $\mathbb{G}(P_{11})$ to every other vertex in $\mathbb G(P)$. Note that the vertex set of  $\mathbb{G}(P_{11})$ is the root vertex set of $\mathbb{G}(P)$. If $P$ is irreducible, then $r=N$.

Without loss of generality, assume that $P$ is in the form of (\ref{eq:Ptree}). We discuss two scenarios when $\mathbb{G}(P)$ contains edges of mixed signs: the first  is that $\mathbb{G}(P_{11})$ is structurally unbalanced and the other is that $\mathbb{G}(P_{11})$ is structurally balanced with at least one negative edge.

\noindent{\it Case 1. $\mathbb{G}(P_{11})$ is structurally unbalanced.}

Since $\mathbb{G}(P_{11})$ is structurally unbalanced, the subgraph with the vertex set $\{v_1^+,\ldots,v_r^+,v_1^-,\ldots,v_r^-\}$ in the graph $\mathbb{G}(Q)$ is strongly connected. In addition, since $\mathbb{G}(P)$ contains a spanning tree, similar to the proof of Lemmas \ref{lm:7} and \ref{lm:8}, one can show that $\mathbb{G}(Q)$ contains a spanning tree as well. $y(t)$ in system (\ref{eq:ysystem2}) converges to $\alpha \bm{1}$ for some constant $\alpha$ as $t\rightarrow \infty$, where $\bm{1}$ is the all-one vector of compatible dimension; in addition $\alpha$ must be 0 since $\lim_{t\rightarrow\infty}(x_i^+(t)+x_i^-(t))=2\alpha=0.$ Thus $x(t)\rightarrow0$ as $t\rightarrow \infty$.

\noindent{\it Case 2. $\mathbb{G}(P_{11})$ is structurally balanced with at least one negative edge.}

If $\mathbb{G}(P)$ is structurally balanced, then similar to Lemma \ref{lm:7} one can show that $\mathbb{G}(Q)$ contains two disconnected components with $v_i^+$ in one component and $v_i^-$ in the other.  In addition, each component contains a spanning tree. Thus the agents in each component reaches the same value which is the opposite of the other component. It immediately implies that the agents in (\ref{eq:system1_2}) polarize.

Next we consider the case when $\mathbb{G}(P)$ is structurally unbalanced. Since $\mathbb G(P_{11})$ is structurally balanced, we know from Lemma \ref{lm:7} that in graph $\mathbb G(Q)$ the subgraph with the vertex set $\bar{\mathcal V}^s=\{v_1^+,\ldots,v_r^+,v_1^-,\ldots,v_r^-\}$ is composed of two disconnected components and each one is strongly connected. The vertex sets of the two components can be denoted as $\bar{\mathcal V}^s_1=\{v_{i_1}^+,\ldots,v_{i_{m}}^+,v_{i_{m+1}}^-,\ldots,v_{i_r}^-\}$ and $\bar{\mathcal V}^s_2=\{v_{i_1}^-,\ldots,v_{i_{m}}^-,v_{i_{m+1}}^+,\ldots,v_{i_r}^+\}$, $1\leq i_1,\ldots,i_r\leq r$. Since $\mathbb G(P)$ contains a spanning tree, one know that for every vertex in $\bar{\mathcal V}\backslash\bar{\mathcal V}^s$ in $\mathbb G(Q)$, there is a directed path from some vertex in $\bar{\mathcal V}^s$ to it.

We check the spectral property of $Q$ and determine the limit of $Q^k$ as $k\rightarrow\infty.$ Using some permutation of rows and columns of $Q$, $Q$ can be transformed into
\begin{equation}\label{eq:Q}
Q=\begin{bmatrix}
         Q_1 & \bm{0} & \bm{0}  \\
         \bm{0}  & Q_2 & \bm{0}  \\
         Q_{31} & Q_{32} & Q_{33} \\
       \end{bmatrix},
       \end{equation}
with $Q_1=Q_2$ and $Q_1$ being irreducible. The vertices in the subgraph $\mathbb G(Q_1)$ are $\bar{\mathcal V}^s_1$ and those in $\mathbb G(Q_2)$ are $\bar{\mathcal V}^s_2$, and for any vertex in $\mathbb G(Q_{33})$ there is a directed path from some vertex either in $\mathbb G(Q_1)$ or in $\mathbb G(Q_2)$ to it. It can be shown that the spectral radius of $Q_{33}$ is less than 1, i.e., $\rho(Q_{33})<1$ \cite{XiWa06}. Since $Q_1$ has positive diagonals and is irreducible, 1 is a simple eigenvalue of $Q_1$ and the magnitudes of all the other eigenvalues of $Q_1$ are less than 1. Hence $Q$ has exactly two eigenvalues equal to 1 and the magnitudes of all the other eigenvalues are less than 1. The following lemma, the proof of which is provided in Appendix \ref{ap:C}, is useful for determining the asymptotic state of system (\ref{eq:ysystem2}).

\begin{lemma}\label{lm:10}
Let $Q=(q_{ij})_{s\times s}=\begin{bmatrix}
         Q_1 & \bm{0} & \bm{0}  \\
         \bm{0}  & Q_1 & \bm{0}  \\
         Q_{31} & Q_{32} & Q_{33} \\
       \end{bmatrix}$ be a stochastic matrix, where $Q_1\in\R^{r\times r}$ is a square matrix. Assume that $Q$ has exactly two eigenvalues equal to 1 and the magnitudes of all the other eigenvalues are less than 1. Then
\begin{equation}\label{eq:lm1_1}
\lim_{k\rightarrow\infty}Q^k=\begin{bmatrix}
         \bm{1}\xi^T & \bm{0}  & \bm{0}  \\
         \bm{0}  & \bm{1}\xi^T  & \bm{0} \\
         \eta_1\xi^T & \eta_2\xi^T & \bm{0} \\
       \end{bmatrix},
\end{equation}
where $\xi\geq0$, $\xi^TQ_1=\xi^T,\ \xi^T\bm{1}=1$ and $\eta_1=(I-Q_{33})^{-1}Q_{31}\bm{1},\eta_2=(I-Q_{33})^{-1}Q_{32}\bm{1}$ are some nonnegative column vectors. In addition, $||\eta_1-\eta_2||_\infty\leq1.$
\end{lemma}

The matrix $Q$ in (\ref{eq:Q}) satisfies the condition in Lemma \ref{lm:10}. Hence the states of all the agents in (\ref{eq:ysystem2}) converge. The agents in  $\bar{\mathcal V}^s$ have the same final absolute value given by $|\xi^T[x_{i_1}(0),\ldots,x_{i_{m}}(0),-x_{i_{m+1}}(0),\ldots,-x_{i_r}(0)]^T|$, where $\xi$ is the left eigenvector of $Q_1$ corresponding to 1 defined as in the above lemma. For every agent in $\mathcal V\backslash\bar{\mathcal V}^s$, we have the following bound
{\small
\begin{align*}
&\lim_{t\rightarrow\infty}|y_i(t)|\\
\leq&||\eta_1-\eta_2||_\infty|\xi^T[x_{i_1}(0),\ldots,x_{i_{m}}(0),-x_{i_{m+1}}(0),\ldots,-x_{i_r}(0)]^T|\\
\leq&|\xi^T[x_{i_1}(0),\ldots,x_{i_{m}}(0),-x_{i_{m+1}}(0),\ldots,-x_{i_r}(0)]^T|,
\end{align*}}
for $i=2r+1,\ldots,2N$.

We summarize the above discussion into the following theorem.

\begin{theorem}\label{thm:tree1}
Consider $P$ in the form of (\ref{eq:Ptree}), with $P_{11}$ being irreducible, $|P|$ being stochastic, and $\mathbb{G}(P)$ containing a spanning tree. If $\mathbb{G}(P_{11})$ is structurally unbalanced, the state of system (\ref{eq:system1_2}) converges to zero for every initial value. If $\mathbb{G}(P_{11})$ is structurally balanced with at least one negative edge, then the agents of the subgraph $\mathbb{G}(P_{11})$ polarize, and the states of the other agents converge and lie in the interval $[-|C|,|C|]$, where $|C|$ is the absolute value of the polarized value of the agents in $\mathbb{G}(P_{11})$. Furthermore, when $\mathbb{G}(P)$ is structurally balanced with at least one negative edge, system  (\ref{eq:system1_2}) polarizes.
\end{theorem}

\begin{remark}
In \cite{HuZh13}, the authors pointed out that when the fixed graph is structurally unbalanced and contains a spanning tree, the states of the agents may converge to zero or become fragmented. Here by looking into the eigenvalues and eigenvectors of system matrix $Q$ of (\ref{eq:ysystem2}),  we are able to give a complete characterization of the final state of system (\ref{eq:system1_2}).
\end{remark}

%=======================================================================
\subsection{$\mathbb G(P(t))$ is time-varying}

In this subsection, we consider the case when $\mathbb G(P(t))$ changes with time. Assume that there exists a constant $\gamma,\  0<\gamma<1,$ such that the nonzero elements of $P(t)$ satisfy
\begin{equation}\label{eq:lowerBound}
|(P(t))_{ij}|\geq\gamma\ \text{for}\ (P(t))_{ij}\neq0,\ t=0,1,2,\ldots.
\end{equation}
We have the following polarization result.
\begin{theorem}\label{thm:3}
Assume that $P(t),\ t=0,1,2,\ldots,$ satisfy (\ref{eq:rowSumT}) and (\ref{eq:lowerBound}) and there exists a bipartition of $\mathcal{V}$ into two nonempty subsets, such that for each graph $\mathbb{G}(P(t))$, $t\geq0$, the edges between the two subsets are negative and the edges within each subset are positive. Assume that there exists an infinite sequence of nonempty, uniformly bounded time intervals $[t_i,t_{i+1})$, $i\geq0$, starting at $t_0=0$ with the property that across each time interval $[t_i,t_{i+1})$, the union of the graphs $\mathbb{G}(P(t))$ is strongly connected. Then system (\ref{eq:system1_2}) polarizes.
\end{theorem}
{\it Proof.} Assume that the bipartition of $\mathcal V$ is $\mathcal{V}_1=\{v_{i_1},\ldots,v_{i_m}\},\ \mathcal{V}_2=\{v_{i_{m+1}},\ldots,v_{i_N}\}$ for some $1\leq m<N$ and $1\leq i_1,\ldots,i_N\leq N$. Consider system (\ref{eq:ysystem2}). From the assumption, one can show that across each time interval $[t_i,t_{i+1})$, the union of the graphs $\mathbb{G}(Q(t))$ contains two disconnected components with the vertex sets $\{v_{i_1}^+,\ldots,v_{i_{m}}^+,v_{i_{m+1}}^-,\ldots,v_{i_N}^-\}$ and $\{v_{i_1}^-,\ldots,v_{i_{m}}^-,v_{i_{m+1}}^+,\ldots,v_{i_N}^+\}$, and in addition, each component is strongly connected. We conclude from \cite{ReBe05,CaMoAn08a} that the states of the agents in each of the two subsystems converge to the same values, respectively, which are the opposite of each other. Since the initial conditions that render the agreed value of each component to be 0 come from a set with zero measure, we know that system (\ref{eq:system1_2}) polarizes.
\hfill $\Box$

\begin{remark}
If $P(t),\ t=0,1,2,\ldots,$ are all irreducible and structurally balanced and furthermore the unique bipartitions of $\mathcal{V}$ satisfying Definition \ref{def:balanced} are the same for $\mathbb{G}(P(t)),\ t=0,1,2,\ldots$, then the assumptions in Theorem \ref{thm:3} are satisfied and the states of the agents converge to two opposite values. Theorem \ref{thm:3} is a generalization of the previous results for distributed averaging algorithms in \cite{ReBe05,CaMoAn08a}, where the weights are nonnegative and obviously $P(t)$ are structurally balanced.
\end{remark}

\begin{theorem}\label{thm:4}
Let $P(t),\ t=0,1,2,\ldots,$ satisfy (\ref{eq:rowSumT}) and (\ref{eq:lowerBound}). Assume that $[t_i,t_{i+1})$, $i\geq0$, $t_0=0$, is an infinite sequence of  nonempty, uniformly bounded time intervals. Suppose that across each time interval $[t_i,t_{i+1})$, the union of the graphs is strongly connected and there does not exist a bipartition of $\mathcal{V}$ into two subsets, such that for each graph $\mathbb{G}(P(s))$, $s\in[t_i,t_{i+1})$, the edges between the two subsets are negative and the edges within each subset are positive. Then $x(t)$ of system (\ref{eq:system1_2}) converges to zero asymptotically.
\end{theorem}

Note that in Theorem \ref{thm:4}, one of the two subsets may be empty.

\begin{remark}
For each time interval $[t_i,t_{i+1})$, if there always exists some $t\in[t_i,t_{i+1})$, such that $P(t)$ is strongly connected and structurally unbalanced, then the conditions in Theorem \ref{thm:4} are satisfied and thus the state of the system converges to zero. Said differently, if structural unbalance arises in the network frequently
enough, then polarization of the states of the agents will not happen and instead the opinions of the agents in the network become neutralized in the end.
\end{remark}

{\it Proof of Theorem \ref{thm:4}.} It suffices to prove that $y(t)$ of system (\ref{eq:ysystem2}) converges to $\alpha\bm{1}$  for some constant $\alpha$ as $t$ goes to infinity. For each time interval $[t_i,t_{i+1})$, we will prove that the union of the graphs over $[t_i,t_{i+1})$, $\cup_{s\in[t_i,t_{i+1})}\mathbb{G}(Q(s))$, is strongly connected. Then from Theorem 3.10 in \cite{ReBe05}, it follows that $y(t)$ converges to $\alpha\bm{1}$ as $t$ goes to infinity.

For each time interval $[t_i,t_{i+1})$, define a directed graph $\mathbb{G}^m=(\mathcal{V}^m,\mathcal{E}^m)$ with $\mathcal{V}^m=\mathcal{V}$ as follows. For two vertices $v_j$ and $v_k$, there exists a positive edge $(v_j,v_k)\in\mathcal{E}^m$ if $(v_j,v_k)$ is a positive edge in graph $\mathbb{G}(P(s))$ for some $s\in[t_i,t_{i+1})$; there is a negative edge $(v_j,v_k)\in\mathcal{E}^m$ if $(v_j,v_k)$ is a negative edge in graph $\mathbb{G}(P(s))$ for some $s\in[t_i,t_{i+1})$. Note that  for an ordered pair of vertices $v_j$ and $v_k$, there may exist two directed edges $(v_j,v_k)$ in $\mathbb{G}^m$ with one being positive and the other being negative. Let the enlarged graph based on $\mathbb{G}^m$ be $\overline{\mathbb{G}^m}$. Since the union of the graphs $\mathbb{G}(P(s))$  over the interval $[t_i,t_{i+1})$ is strongly connected,  $\mathbb{G}^m$ is strongly connected. In addition, from the condition that there does not exist a bipartition of $\mathcal{V}$ into two subsets, such that for each graph $\mathbb{G}(P(s))$, $s\in[t_i,t_{i+1})$, the edges between the two subsets are negative and the edges within each subset are positive, there is a negative cycle in the graph $\mathbb{G}^m$. Mimicking the proof in the necessity part of Lemma~\ref{lm:8}, it can be proved that the enlarged graph $\overline{\mathbb{G}^m}$ is strongly connected. Based on the way we define $\overline{\mathbb{G}^m}$, it can be seen that $\overline{\mathbb{G}^m}$ and $\cup_{s\in[t_i,t_{i+1})}\mathbb{G}(Q(s))$ are isomorphic and thus $\cup_{s\in[t_i,t_{i+1})}\mathbb{G}(Q(s))$ is strongly connected. Hence, $y(t)$ converges to $\alpha\bm{1}$ asymptotically, which implies the state $x(t)$ converges to zero asymptotically.   \hfill $\Box$

If the union of the graphs over $[t_i,t_{i+1})$, $\cup_{s\in[t_i,t_{i+1})}\mathbb{G}(P(s))$, is not strongly connected, but only contains a spanning tree, system (\ref{eq:system1_2}) can give rise to some new behavior as discussed next\footnote{We are indebted to Prof. M. Elena Valcher for pointing out a missing assumption in the journal version \cite{XiCaJo16} of Theorem \ref{thm:5}.}.

\begin{theorem}\label{thm:5}
Let $P(t),\ t=0,1,2,\ldots,$ satisfy (\ref{eq:rowSumT}) and (\ref{eq:lowerBound}). Let the root vertex set of the union of the graphs over $[0,\infty)$ be $\mathcal{V}^s$. Assume that there exists a bipartition of $\mathcal{V}^s$ into two nonempty subsets, such that for each graph $\mathbb{G}(P(t))$, $t\geq0$, the edges between the two subsets are negative and the edges within each subset are positive. Assume that there exists an infinite sequence of  nonempty, uniformly bounded time intervals $[t_i,t_{i+1})$, $i\geq0$, starting at $t_0=0$ with the property that across each time interval $[t_i,t_{i+1})$, the union of the graphs contains a spanning tree and the
root vertex set of the union of the graphs is $\mathcal{V}^s$. Then the agents in the root vertex set polarize, and the other agents' states will finally lie in between the polarized values.
\end{theorem}
{\it Proof.} From Theorem \ref{thm:3} we know that the agents in the root vertex set $\mathcal V^s$ polarize. Assume that the polarized values are $C$ and $-C$, where $C$ is a nonnegative constant. We prove that the states of the other agents will asymptotically be bounded by $C$.

Let $$C(t)=\max_{v_i\in\mathcal V^s}|x_i(t)|,\ M(t)=\max_{v_i\in\mathcal V\backslash \mathcal V^s}|x_i(t)|$$ for $t\geq0$. From the result in the previous paragraph, one knows that $\lim_{t\rightarrow\infty}C(t)=C$. If $M(t)>C(t)$ holds only for a finite time $t$, then there exists a $t^\prime$ such that $M(t)\leq C(t)$ for $t\geq t^\prime$ and hence
$$\limsup_{t\rightarrow\infty}M(t)\leq\lim_{t\rightarrow\infty}C(t)=C.$$
For this case, the desired conclusion follows.

Next we assume that $M(t)>C(t)$ holds for an infinite time sequence $t=t^\ast_1,t^\ast_2,\ldots.$ We pick the specific time $t_1^\ast$ to carry out the discussion. It is easy to see from (\ref{eq:rowSumT}) and (\ref{eq:system1_2}) that
\begin{equation}\label{eq:thm5_2}
C(t^\ast_1+l)\leq C(t^\ast_1)<M(t^\ast_1),\ M(t^\ast_1+l)\leq M(t^\ast_1),
\end{equation}
for all $l\geq0$. Pick an integer $r$ such that $t_{r-1}\leq t^\ast_1< t_r$ and consider the time interval $[t_r,t_{r+1})$. Since the union of the graphs across $[t_r,t_{r+1})$ contains a spanning tree, there exists some time $s_1\in[t_r,t_{r+1})$ such that $(v_{i_0},v_{i_1})$ is an edge of the graph $\mathbb G(P(s_1))$ with $v_{i_0}\in\mathcal V^s$ and $v_{i_1}\in\mathcal V\backslash\mathcal V^s$. One has
\begin{eqnarray*}
&&|x_{i_1}(s_1+1)|=\left|\sum_{j=1}^Np_{i_1j}(s_1)x_j(s_1)\right|\\
&\leq&|p_{i_1i_0}(s_1)x_{i_0}(s_1)|+\sum_{j\neq i_0}|p_{i_1j}(s_1)x_j(s_1)|\\
&\leq&\gamma C(t^\ast_1)+(1-\gamma)M(t^\ast_1)\\
&=&C(t^\ast_1)+(1-\gamma)(M(t^\ast_1)-C(t^\ast_1)),
\end{eqnarray*}
where $\gamma$ is the constant in (\ref{eq:lowerBound}). Since $P(t)$ has positive diagonals, further calculation shows that
\begin{eqnarray*}
&&|x_{i_1}(s_1+2)|\\
&\leq&\gamma (C(t^\ast_1)+(1-\gamma)(M(t^\ast_1)-C(t^\ast_1)))+(1-\gamma)M(t^\ast_1)\\
&=&C(t^\ast_1)+(1-\gamma^2)(M(t^\ast_1)-C(t^\ast_1)).
\end{eqnarray*}

Recursively, for $l\geq0$,
\begin{eqnarray*}
|x_{i_1}(s_1+l)|\leq C(t^\ast_1)+(1-\gamma^l)(M(t^\ast_1)-C(t^\ast_1)).
\end{eqnarray*}
Specifically, the following inequality is true for $l\geq0$
\begin{align}\label{eq:thm5_1}
|x_{i_1}(t_{r+1}+l)|&\leq C(t^\ast_1)+(1-\gamma^{t_{r+1}-s_1+l})(M(t^\ast_1)-C(t^\ast_1))\nonumber\\
&\leq C(t^\ast_1)+(1-\gamma^{t_{r+1}-t_r+l})(M(t^\ast_1)-C(t^\ast_1)).
\end{align}
Define $\mathcal V_1=\{v_j| (v_{i},v_{j})$  as an edge in the union of the graphs across the  time interval $[t_r,t_{r+1})$ for some $v_{i}\in\mathcal V^s$ and $v_j\in\mathcal V\backslash\mathcal V^s\}$. Then the above inequality (\ref{eq:thm5_1}) holds for any $v_{i_1}\in\mathcal V_1$.

Consider the time interval $[t_{r+1},t_{r+2}).$ Define $\mathcal V_2=\{v_j| (v_{i},v_{j})$ as an edge in the union of the graphs across the  time interval $[t_{r+1},t_{r+2})$ for some $v_{i}\in\mathcal V^s\cup\mathcal V_1$ and $v_j\in\mathcal V\backslash(\mathcal V^s\cup\mathcal V_1)\}$. If $v_{i_2}\in\mathcal V_2$, $v_{i_0}\in\mathcal V^s\cup\mathcal V_1$ and $(v_{i_0},v_{i_2})$  is an edge of the graph $\mathbb{G}(P(s_2))$ for some $s_2\in[t_{r+1},t_{r+2})$, one has
\begin{equation*}
\begin{split}
&|x_{i_2}(s_2+1)|\\
\leq&|p_{i_2i_0}(s_2)x_{i_0}(s_2)|+\sum_{j\neq i_0}|p_{i_2j}(s_2)x_j(s_2)|\\
\leq&\gamma (C(t^\ast_1)+(1-\gamma^{s_2-t_r})(M(t^\ast_1)-C(t^\ast_1)))+(1-\gamma)M(t^\ast_1)\\
=&C(t^\ast_1)+(1-\gamma^{s_2-t_r+1})(M(t^\ast_1)-C(t^\ast_1)).
\end{split}
\end{equation*}
Thus for $l\geq0$, the following inequality holds
\begin{eqnarray*}
|x_{i_2}(s_2+l)|\leq C(t^\ast_1)+(1-\gamma^{s_2-t_r+l})(M(t^\ast_1)-C(t^\ast_1)).
\end{eqnarray*}
For all $v_i\in\mathcal V_1\cup\mathcal V_2$, it holds that
\begin{eqnarray*}
|x_{i}(t_{r+2}+l)|\leq C(t^\ast_1)+(1-\gamma^{t_{r+2}-t_r+l})(M(t^\ast_1)-C(t^\ast_1)).
\end{eqnarray*}

Continuing this process, one derives that for all $v_i\in\mathcal V\backslash\mathcal V_r$,
\begin{equation*}
\begin{split}
|x_{i}(t_{r+N-1})|
&\leq C(t^\ast_1)+(1-\gamma^{t_{r+N-1}-t_r})(M(t^\ast_1)-C(t^\ast_1))\\
&\leq C(t^\ast_1)+(1-\gamma^{(N-1)T})(M(t^\ast_1)-C(t^\ast_1)),
\end{split}
\end{equation*}
where $T$ is a uniform upper bound for $t_{r+1}-t_r$. Repeating the above calculation, we have that for all $v_i\in\mathcal V\backslash\mathcal V_r$,
\begin{eqnarray*}
|x_{i}(t_{r+(N-1)l})|\leq C(t^\ast_1)+(1-\gamma^{(N-1)T})^l(M(t^\ast_1)-C(t^\ast_1)).
\end{eqnarray*}
Combining with (\ref{eq:thm5_2}), one has that
\begin{align*}
&M(t_{r+(N-1)l}+s)\leq \max\{M(t_{r+(N-1)l}),C(t_{r+(N-1)l})\}\\
\leq& C(t^\ast_1)+(1-\gamma^{(N-1)T})^l(M(t^\ast_1)-C(t^\ast_1)),
\end{align*}
for all $0\leq s<t_{r+(N-1)(l+1)}-t_{r+(N-1)l}$. From the above inequality, we can conclude that $\limsup\limits_{t\rightarrow\infty}M(t)\leq C(t^\ast_1)$. Since the above discussion applies to all $t^\ast_r$, it holds that $\limsup\limits_{t\rightarrow\infty}M(t)\leq C(t^\ast_r)$ for all $r=1,2,\ldots.$ In view of the fact that $\lim_{t\rightarrow \infty}C(t)=C$, one has $\limsup\limits_{t\rightarrow\infty}M(t)\leq C.$ This completes the proof.
 \hfill $\Box$

\begin{remark}
In the fixed topology case in the previous section, when the graph contains a spanning tree, it is shown in Theorem \ref{thm:tree1} that  the agents in the root vertex set polarize and the states of the other agents converge and lie in between the polarized values. However, when the network topologies are time-varying, the states of the other agents may not converge but they will finally lie in between the polarized values, which will be illustrated though an example in Section~\ref{se:simulation}.
\end{remark}

\begin{theorem}\label{thm:6}
Let $P(t),\ t=0,1,2,\ldots,$ satisfy (\ref{eq:rowSumT}) and (\ref{eq:lowerBound}). Assume that $[t_i,t_{i+1})$, $i\geq0$, $t_0=0$, is an infinite sequence of nonempty, uniformly bounded time intervals. Suppose that across each time interval $[t_i,t_{i+1})$, the union of the graphs contains a spanning tree and for the root vertex set of the union graph, there does not exist a bipartition of this set into two subsets, such that for each graph $\mathbb{G}(P(s))$, $s\in[t_i,t_{i+1})$, the edges between the two subsets are negative and the edges within each subset are positive. Then the state of system (\ref{eq:system1_2}) converges to zero asymptotically.
\end{theorem}
{\it Proof.} Using similar arguments to the proof of Theorem \ref{thm:4}, we can show that the union of the graphs $\mathbb{G}(Q(t))$ across each time interval $[t_i,t_{i+1})$, $i\geq0$, contains a spanning tree. It immediately follows that the $y$-system (\ref{eq:ysystem2}) converges to $\alpha\bm{1}$ for some constant $\alpha$ from \cite{ReBe05}. Thus system (\ref{eq:system1_2}) converges to zero asymptotically.  \hfill $\Box$

\begin{remark}
In \cite{He14}, Hendrickx formally introduced the transformation (\ref{eq:ysystem1}) and studied discrete-time and continuous-time systems with reciprocal interactions between agents and nonreciprocal interactions under joint strong connectivity conditions. The convergence of the system to polarized values or to zero were derived based on studies on consensus systems with ``type-symmetric" interactions and by looking into the persistent interactions between agents \cite{HeTs13}. Here we have considered nonreciprocal interactions between agents with joint graphs containing spanning trees, where opinion separation of the agents may appear.
\end{remark}

%=======================================================================
\section{Continuous-time model}\label{se:continuous}
In this section, we present our main results for the continuous-time model (\ref{eq:system2_2}). For each $A(t)$, similar to (\ref{eq:P}), we can define two nonnegative matrices $A^+(t)$ and $A^-(t)$ based on $A(t)$. Let  $x^+_i(t)=x_i(t)$, $x^-_i(t)=-x_i(t)$ and $y(t)=[x^+_1(t),\ldots,x^+_N(t),x^-_1(t),\ldots,x^-_N(t)]^T$. From system (\ref{eq:system2_2}), we obtain the following update equations for $y(t)$:
\begin{equation}\label{eq:ysystem3}
{\footnotesize \dot{y}(t)=\left(\begin{bmatrix}
      A^+(t) & A^-(t)  \\
      A^-(t)  & A^+(t)  \\
    \end{bmatrix}-\begin{bmatrix}
      D(t) & \bm{0}  \\
      \bm{0}  & D(t)  \\
    \end{bmatrix}\right)y(t)=-W(t)y(t),}
\end{equation}
where $D(t)=\diag\{d_1(t),\ldots,d_N(t)\}$ with $d_i(t)=\sum_{j=1,\ j\neq i}^N|a_{ij}(t)|$ and $W(t)=\begin{bmatrix}
      D(t) & \bm{0}  \\
      \bm{0}  & D(t)  \\
    \end{bmatrix}-\begin{bmatrix}
      A^+(t) & A^-(t)  \\
      A^-(t)  & A^+(t)  \\
    \end{bmatrix}$ is the Laplacian matrix with nonpositive off-diagonal elements. The dynamical behavior of system (\ref{eq:system2_2}) can be revealed by studying system (\ref{eq:ysystem3}).

\subsection{$\mathbb G(L(t))$ is fixed}

Consider the continuous-time system (\ref{eq:system2_2}) under fixed topologies. Let $A(t)\equiv A\in\R^{N\times N}$ be the signed adjacency matrix, and let  $L(t)\equiv L$ be the signed Laplacian matrix given by (\ref{eq:signedL}) for all $t\geq0$. When the graph $\mathbb{G}(L)$ contains a spanning tree, by a suitable permutation of rows and columns of its associated signed Laplacian matrix $L$, $L$ can be brought into the following form
\begin{equation}\label{eq:Ltree}
L=\begin{bmatrix}
    L_{11} & \bm{0} \\
    L_{21} & L_{22} \\
  \end{bmatrix}
  \end{equation}
where $L_{11}\in\R^{r\times r}$ is irreducible, $L_{22}\in\R^{(N-r)\times(N-r)},$ and $L_{21}\in\R^{(N-r)\times r}$. By looking into system (\ref{eq:ysystem3}) and in view of Lemmas~\ref{lm:7} and \ref{lm:8}, similar to Theorem \ref{thm:tree1}, we have the following theorem.

\begin{theorem}\label{thm:tree2}
Let $\mathbb{G}(L)$ be a signed graph containing a spanning tree and let $L$ be its associated signed Laplacian matrix in the form (\ref{eq:Ltree}). If the subgraph $\mathbb{G}(L_{11})$ is structurally unbalanced, the state of system (\ref{eq:system2_2}) converges to zero for every initial value. If $\mathbb{G}(L_{11})$ is structurally balanced with at least one negative edge, then the agents in the subgraph $\mathbb{G}(L_{11})$ polarize and the states of the other agents converge and lie in between the polarized values; furthermore, if $\mathbb{G}(L)$ is structurally balanced with at least one negative edge, system  (\ref{eq:system2_2}) polarizes.
\end{theorem}

%=======================================================================
\subsection{$\mathbb G(L(t))$ is time-varying}

In this subsection, we consider the case when the interaction graph topologies are dynamically changing. Assume that $A(t)$ and $L(t)$ are piecewise constant functions and the interaction graph topologies or the weights of the edges change at time instants $t_1,t_2,\ldots$. System (\ref{eq:system2_2}) can be rewritten as
\begin{equation} \label{eq:system4_3}
\dot{x}(t)=-L(t_i)x(t),\quad t\in[t_i,t_i+\tau_i),
\end{equation}
where $t_0=0$ is the initial time, and $\tau_i=t_{i+1}-t_i,\ i=0,1,\ldots$ are the dwell times. Let $\tau$ be a finite
set of positive numbers and let $\mathcal{T}$ be an infinite set generated from $\tau$, which is closed under addition, and multiplication by positive integers. Assume that $\tau_i\in \mathcal{T}$, $i=0,1,2,\ldots.$ Let the nonzero elements $a_{jk}(t_i)$ of $A(t_i)$ satisfy that $a_{jk}(t_i)\in[\gamma_1,\gamma_2],$ where $\gamma_1,\gamma_2$ are positive constants.

The $y$-system (\ref{eq:ysystem3}) can be written as
\begin{equation}\label{eq:ysystem4}
\dot{y}(t)=-W(t_i)y(t),\quad t\in[t_i,t_i+\tau_i).
\end{equation}
Employing similar ideas as in the previous section for the discrete-time model (\ref{eq:system1_2}) and (\ref{eq:ysystem2}) and in view of Theorem 3.12 in \cite{ReBe05}, we can prove the following two theorems.

\begin{theorem}\label{thm:7}
Let the root vertex set of the union of the graphs over $[0,\infty)$ be $\mathcal{V}^s$. Assume that there exists a bipartition of $\mathcal{V}^s$ into two nonempty subsets, such that for each graph $\mathbb{G}(L(t))$, $t\geq0$, the edges between the two subsets are negative and the edges within each subset are positive. Assume that there exists an infinite sequence of  nonempty, uniformly bounded time intervals $[t_{i_k},t_{i_{k+1}})$, $k\geq0$, starting at $t_{i_0}=0$ with the property that across each time interval $[t_{i_k},t_{i_{k+1}})$, the union of the graphs contains a spanning tree and the root vertex set of the union of the graphs is $\mathcal V^s$. Then the agents in the root vertex set of system (\ref{eq:system2_2}) polarize, and the states of the other agents will finally lie in between the polarized values.
\end{theorem}

\begin{theorem}\label{thm:8}
Assume that $[t_{i_k},t_{i_{k+1}})$, $k\geq0$, $t_{i_0}=0$, is an infinite sequence of nonempty, uniformly bounded time intervals. Suppose that across each time interval $[t_{i_k},t_{i_{k+1}})$, the union of the graphs contains a spanning tree and for the root vertex set of the union graph, there does not exist a bipartition of this set into two subsets, such that for each graph $\mathbb{G}(L(s))$, $s\in[t_{i_k},t_{i_{k+1}})$, the edges between the two subsets are negative and the edges within each subset are positive. Then the state of system (\ref{eq:system2_2}) converges to zero asymptotically.
\end{theorem}
%=======================================================================

\section{Illustrative examples}\label{se:simulation}

\begin{figure}
\begin{center} \includegraphics[width=1.6in]{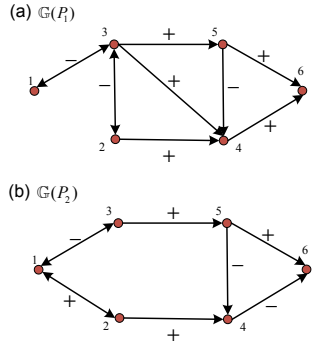}
\caption{Two graphs $\mathbb{G}(P_1)$ and $\mathbb{G}(P_2)$ both contain spanning trees.} \label{fig:ex4_1}
\end{center}
\end{figure}

In this section, we  perform simulation studies on system (\ref{eq:system1_2}) with topologies containing spanning trees. Consider the two graphs shown in Fig. \ref{fig:ex4_1}, where the edges with negative weights are labeled by ``$-$" signs and those with positive weights are labeled by ``$+$" signs. Their corresponding matrices $P_1$ and $P_2$ are given by
\begin{align*}
&P_1=\begin{bmatrix}
     (P_1)_{11} & \bm{0} \\
     (P_1)_{21} & (P_1)_{22} \\
   \end{bmatrix}
=\left[
  \begin{array}{ccc:ccc}
      \frac{1}{2} &0 & -\frac{1}{2} & 0 & 0 &0 \\
     0 & \frac{1}{2} & -\frac{1}{2} & 0 & 0  & 0\\
      -\frac{1}{3} & -\frac{1}{3} & \frac{1}{3} & 0 & 0 & 0\\
                   \hdashline %
      0 & \frac{1}{4} &  \frac{1}{4} & \frac{1}{4} & -\frac{1}{4} & 0\\
      0 & 0 & \frac{1}{2} & 0 & \frac{1}{2} & 0\\
      0 & 0 & 0 & \frac{1}{3} &  \frac{1}{3} & \frac{1}{3} \\
  \end{array}
\right],\\
&P_2=\begin{bmatrix}
     (P_2)_{11} & \bm{0} \\
     (P_2)_{21} & (P_2)_{22} \\
   \end{bmatrix}
=\left[
  \begin{array}{ccc:ccc}
      \frac{1}{3} &\frac{1}{3} & -\frac{1}{3} & 0 & 0 &0 \\
     \frac{1}{2} & \frac{1}{2} & 0 & 0 & 0  & 0\\
      -\frac{1}{2} & 0 & \frac{1}{2} & 0 & 0 & 0\\
                   \hdashline %
      0 & \frac{1}{3} &  0 & \frac{1}{3} & -\frac{1}{3} & 0\\
      0 & 0 & \frac{1}{2} & 0 & \frac{1}{2} & 0\\
      0 & 0 & 0 & -\frac{1}{3} &  \frac{1}{3} & \frac{1}{3} \\
  \end{array}
\right].
\end{align*}
One can see that $\mathbb{G}(P_1)$ is structurally unbalanced but the subgraph $\mathbb{G}((P_1)_{11})$ is structurally balanced. $\mathbb{G}(P_2)$ is structurally balanced. Let the initial state of the system be $x(0)=[0.9, 0.7, -0.9, -1, 0.2, 0.9]^T$.

\begin{figure}
\begin{center} \includegraphics[width=2.6in]{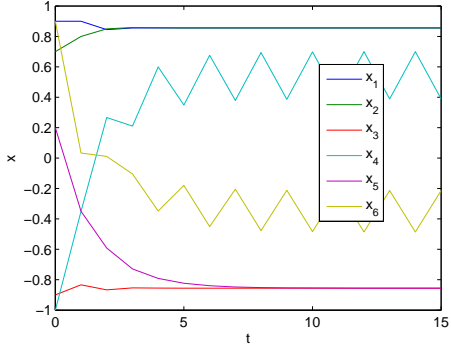}
\caption{The evolution of the agents' states with the graphs switching between Fig. \ref{fig:ex4_1}(a) and Fig. \ref{fig:ex4_1}(b).} \label{fig:ex4_2}
\end{center}
\end{figure}

The evolution of the states of the agents under the graph topology $\mathbb{G}(P_1)$ in Fig. \ref{fig:ex4_1}(a) has been illustrated in Fig. \ref{fig:ex1_2}(b) in Section~\ref{se:mot}. As indicated in Theorem \ref{thm:tree1}, the agents $1,2,3$ in the subgraph $\mathbb{G}(P_{11})$ achieve opposite values and the states of agents 4, 5 and 6 converge and lie in between the opposite values. For system (\ref{eq:system1_2}) with
\begin{eqnarray}\label{eq:switch}
P(t)=\begin{cases} P_1,\  t\ \text{is even},\\
P_2,\   t\ \text{is odd},
\end{cases}
\end{eqnarray} the evolution of the states of the agents are shown in Fig.~\ref{fig:ex4_2}, from which we can see that the states of agents 4 and 6 do not converge but they still lie in between the opposite values of agents $1,2,3$.

%=======================================================================
\section{Conclusion}\label{se:conclusion}
In this paper, we have studied the relationship between structural balance and opinion separation in social networks that contain both trust and mistrust relationships. When the opinion update rules are described by DeGroot-type models, we have shown that under conditions that are closely related to whether a network is structurally balanced or not, the opinions sometimes get separated, for which in the extreme case the network evolves into two polarized  camps, and sometimes become neutralized. Our results complement the existing results in the literature.

We are interested in further developing opinion separation models that rely less on the DeGroot averaging rules. One promising direction is to look into the biased assimilation behavior in social groups. The nonlinearity inherently associated with such behavior is a main challenge that we want to attack.

%=======================================================================

%\bibliographystyle{unsrt}
%\bibliography{e:/ref_ming}

%=======================================================================
\appendices

\section{}\label{ap:C}
\noindent{\it Proof of Lemma \ref{lm:10}} Since $Q_1$ is a stochastic matrix, 1 is an eigenvalue of $Q_1$ with the corresponding eigenvector $\bm{1}$. From the assumption of the lemma, we know that 1 is a simple eigenvalue of $Q_1$ and the magnitudes of all the other eigenvalues of $Q_1$ are less than 1. In addition $\rho(Q_{33})<1$. Thus from the Perron-Frobenius theorem \cite{HoJo85},
$$\lim_{k\rightarrow\infty}Q_1^k=\bm{1}\xi^T,$$
where $\xi\geq0,\xi^TQ_1=\xi^T,$ and $\xi^T\bm{1}=1$.

It is easy to see that $v^T_1=[\xi^T\ \bm{0}^T\ \bm{0}^T]$ and $v^T_2=[ \bm{0}^T\ \xi^T\ \bm{0}^T]$ are two independent left eigenvectors of $Q$ corresponding to 1. One can verify that $u_1=[\bm{1}^T\ \bm{0}^T\ \eta_1^T]^T$ and $u_2=[\bm{0}^T\ \bm{1}^T\ \eta_2^T]^T$ are two independent right eigenvectors of $Q$  corresponding to 1, where $\eta_1$ and $\eta_2$ are given by
$$\eta_1=(I-Q_{33})^{-1}Q_{31}\bm{1},\ \eta_2=(I-Q_{33})^{-1}Q_{32}\bm{1}.$$
$I-Q_{33}$ is invertible because $\rho(Q_{33})<1$. In addition, the following equalities hold $v^T_1u_1=1,\ v^T_2u_2=1,\ v^T_1u_2=0$ and $v^T_2u_1=0$.  By using the Jordan canonical form, we can show that $Q^k$ converges as $k$ goes to infinity and
\begin{align*}
\lim_{k\rightarrow\infty}Q^k=u_1v^T_1+u_2v^T_2=\begin{bmatrix}
         \bm{1}\xi^T & \bm{0}  & \bm{0}  \\
         \bm{0}  & \bm{1}\xi^T  & \bm{0} \\
         \eta_1\xi^T & \eta_2\xi^T & \bm{0} \\
       \end{bmatrix}.
\end{align*}

Since $\bm{1}=(Q_{31}+Q_{32}+Q_{33})\bm{1}$, it follows that $\eta_1+\eta_2=(I-Q_{33})^{-1}(Q_{31}+Q_{32})\bm{1}=\bm{1}$. From the nonnegativity of the vectors $\eta_1$ and $\eta_2$, one has that
$$||\eta_1-\eta_2||_\infty\leq||\eta_1+\eta_2||_\infty=1.$$
\hfill $\Box$

\end{document}